\documentclass[review]{elsarticle}
\usepackage{multirow}
\usepackage{hyperref}
\usepackage{amsmath}
\usepackage{amsfonts}
\usepackage{soul,color}
\soulregister\cite7
\soulregister\ref7
\soulregister\pageref7
\journal{Computerized Medical Imaging and Graphics}

\usepackage{numcompress}

\begin{document}

\begin{frontmatter}

\title{Memory-efficient GAN-based Domain Translation of High Resolution 3D Medical Images\footnote{©2020. This manuscript version is made available under the CC-BY-NC-ND 4.0 license http://creativecommons.org/licenses/by-nc-nd/4.0/}}


\author{Hristina~Uzunova}
\cortext[mycorrespondingauthor]{Corresponding author}
\ead{uzunova@imi.uni-luebeck.de}

\author{Jan Ehrhardt, and Heinz Handels}
\address{Institute of Medical Informatics, University of L\"ubeck\\ Ratzeburger Allee 160, L\"ubeck, Germany}

\begin{abstract}

 Generative adversarial networks (GANs) are currently rarely applied on 3D medical images of large size, due to their immense computational demand. The present work proposes a multi-scale patch-based GAN approach for establishing unpaired domain translation by generating 3D medical image volumes of high resolution in a memory-efficient way.

The key idea to enable memory-efficient image generation is to first generate a low-resolution version of the image followed by the generation of patches of constant sizes but successively growing resolutions. To avoid patch artifacts and incorporate global information, the patch generation is conditioned on patches from previous resolution scales. Those multi-scale GANs are trained to generate realistically looking images from image sketches in order to perform an unpaired domain translation. This allows to preserve the topology of the test data and generate the appearance of the training domain data. 

 The evaluation of the domain translation scenarios is performed on brain MRIs of size $\mathbf{155\times240\times240}$ and thorax CTs of size up to $\mathbf{512^3}$.  Compared to common patch-based approaches, the multi-resolution scheme enables better image quality and prevents patch artifacts. Also, it ensures constant GPU memory demand independent from the image size, allowing for the generation of arbitrarily large images.  

\end{abstract}


\end{frontmatter}


\section{Introduction}
Generative adversarial networks (GANs)~\cite{GANs} have shown to be well suited for the generation of photo-realistic images \cite{pix2pixHD,stylegan}. Many medical image analysis and processing applications could also benefit from artificial datasets of realistic images, e.g. by generating ground truth data for augmentation purposes~\cite{trainingcnns,liveraug}, image reconstruction~\cite{dagan}, or domain translation~\cite{mrtransl}. 
The generation of high-resolution images using GANs is, however, still a significant hurdle.
Especially in recent years, the development of GANs towards generation of high-resolution images has been progressing rapidly. In~\cite{progressive}, Karras et al. developed a training procedure for GANs that starts with a low resolution and progressively adds more and more details until the highest resolution level is reached. In this way, they are able to generate highly detailed $1024\times 1024$  images. In~\cite{pix2pixHD} even a larger size of $2048\times 1024$ was reached by using one network to generate low-resolution images and a second one to increase the resolution. However, those training methods already require 16 respectively 24 GB of GPU RAM, which indicates that larger images would require special and expensive hardware. Thus, despite the impressive results achieved on 2D photo-realistic images, the task of generating large 3D images is still rather complex. However, most medical image applications use large-scale 3D images such as thorax CTs and brain MRIs, hence due to  size limitations, GANs are currently impractical for medical applications. In~\cite{3DMedGAN1}, Shin et al. claimed to be forced to use only the half of the image size ($128\times128\times 54$) due to memory restrictions, even though dedicated hardware (NVIDIA DGX system) was used. The largest 3D output size of a GAN found in literature is $128^3$~\cite{3DMedGAN2}, which is far from the actual image size of many medical datasets (e.g. BRATS~\cite{brats}: $155\times240\times240$, LPBA40~\cite{lpba40}: $181\times217\times217$,  COPDgene~\cite{copdgene}: $512 \times512 \times >100$, VISCERAL \cite{visceral}: $>800\times512\times512$ ). A common approach to overcome the computational restraints is the patch-/slice-based generation of images~\cite{SPIE3DGAN,superresolution}. However, when patches or slices are generated independently, artifacts appear on the non-continuous transitions between them. Usually, applying patch overlaps and averaging the values in the overlapping regions would help prevent artifacts, but leads to blurry results. 

An intuitive idea to prevent inconsistencies would be to introduce more global intensity information to the patches. Thus, \cite{deepmedic} proposed to additionally observe a larger area around each patch of the input image to cope with this issue. 
Even though this approach is shown to be well suitable for segmentation and would probably improve patch artifacts in strictly paired image translation (e.g. CT to MR), it cannot be applied to image generation from scratch (or sparsely conditioned), and its effect is limited when the image size drastically exceeds the chosen patch size.

In this work, we propose a memory-efficient multi-scale GAN approach for the generation of large high-resolution 3D medical images. Here, the advantages of a multi-scale approach and patch-based image generation are combined. Our method is designed to first learn a low-resolution version of the whole image. In a second step, patches of higher resolutions are generated in a multi-scale manner, achieved by conditioning the patches of each current scale on patches of the previous level. In this way global information from previous scales is propagated up, and no artifacts appear. Moreover, by only learning a simple task in each step, it is possible to achieve images of higher resolutions.

\begin{figure}[t]
   \centering
    \includegraphics[width=0.6\textwidth]{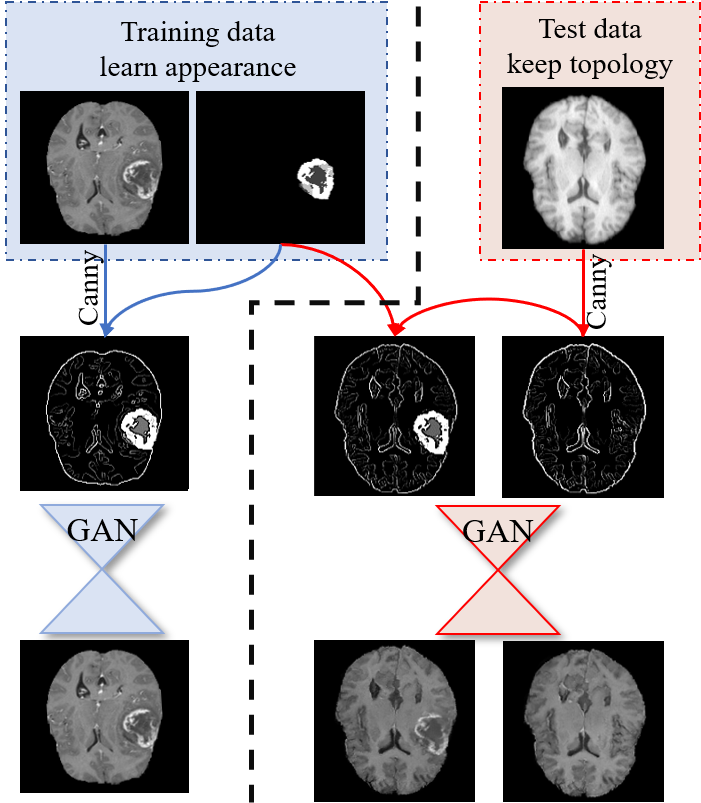}
     \caption{Unpaired domain translation. The extracted edges reassure that the topology of an arbitrary image is preserved and the training dataset defines the output appearance. Also it is possible to explicitly model pathologies.}
     \label{fig:edgesoverview}
\end{figure}

The proposed GAN-based method is applied in an unpaired domain translation use-case scenario. The key idea is similar to \cite{pix2pix}, where the edges of the training images are extracted and used as an input of the GAN. Thus in the training phase the GAN learns to generate a realistically looking appearance where the topology of the input edges is preserved. In test phase, this results into keeping the topology of any source edge image, but adopting the appearance of the training data domain (Fig.~\ref{fig:edgesoverview}). An additional advantage features the possibility to explicitly integrate and model pathologies as shown in Fig.~\ref{fig:edgesoverview}.

The abilities of our method are demonstrated on different datasets, including 3D thorax CTs of sizes up to $512^3$ and brain MRIs of sizes up to $155\times240\times 240$. Furthermore, we show that the presented method has a constant memory demand with growing side length of an isotropic 3D image, whereas other popular methods have an at least cubic growth of the memory demand. Our application features unpaired image domain translation. This is achieved by additionally conditioning the GANs on the image edges to ensure topology preservation of the source image, but allow appearance transfer from the target training data. Unlike conventional paired methods, this approach does not require corresponding images from two domains for training and inference and enables the translation from arbitrary domains to a target domain. To underline those properties, our experiments do not only show standard domain translation tasks ( e.g. between different MRI sequences or CT images reconstructed with different kernels), but also domain translation of data acquired with completely different devices, settings, and patient populations. 

Compared to the preliminary version of this work presented at a conference \cite{uzunovamiccai2019}, herein the following  contributions are made: we include a detailed description of the architecture, training and augmentation strategy as well as extended discussions. Furthermore, we improved the network such that the patches at higher scales consider more contextual information from the lower scales. This results in even more realistic images. Additional experiments investigate the influence of edge and multi-scale information in our approach. We also demonstrate the ability of our method to be used for advanced data augmentation techniques by explicitly integrating varying pathologies into the generated images, thus underlining the benefits of unpaired domain translation. For the first time, experiments for unsupervised domain translation of brain MRIs are included and a quantitative comparison to state-of-the-art paired domain translation methods confirm the quality of the generated images.

This work is organized as follows: After the introduction section, sec. \ref{sec:methods} contains a theoretical description of the proposed multi-scale patch-based GAN method, followed by concrete implementation details. In the next section the experiments and results are presented, starting with experiments showing the need for memory efficient methods for medical images. In sec. \ref{sec:domaintranslation} domain translation experiments on lung CT and brain MRI data are introduced, where both quantitative and qualitative results are shown. At the end, the work is summarized and discussed in sec. \ref{sec:conclusion}.

\section{Methods}
\label{sec:methods}
\begin{figure*}[t]
   \centering
     \includegraphics[width=0.99\textwidth]{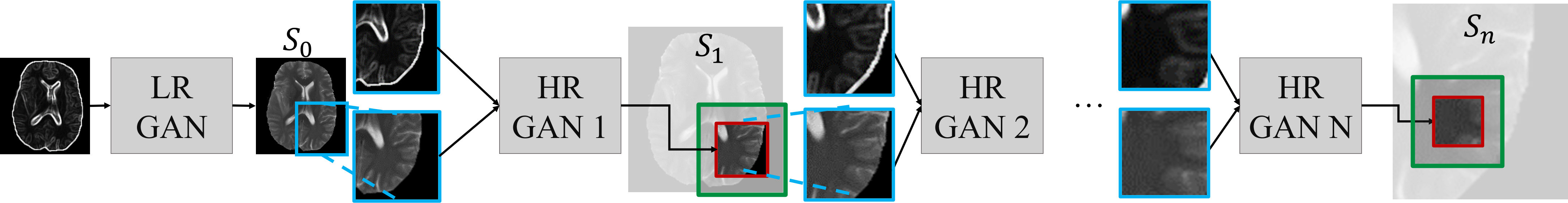}
     \caption{An overview of the proposed GAN. Generate the whole image with a low resolution (LR) GAN, then subsequently increase the resolution by generating patches with multiple high resolution (HR) GANs conditioned on the previous scales.  Blue: an input patch at the current resolution scale; red: a generated patch of a subarea of the green patch at the next higher resolution scale; green: the reception field of the current generated red patch.}
     \label{fig:overview}
\end{figure*}
GANs are generative models able to generate realistic images of higher quality compared to other generative methods such as (variational) autoencoders \cite{pix2pix}. GANs learn to map a random noise vector $\mathbf{z}$ to an output image $y$ using a generator function $G: \mathbf{z}\rightarrow y$ ~\cite{GANs}. To ensure that the generator produces realistically looking image ($y$) that cannot be distinguished from real ones ($\tilde y$), an adversarial discriminator $D$ is enclosed in the training process, aiming to perfectly distinguish between real images and generator's fakes. In an alternating manner, the generator tries to fool the discriminator by generating more realistic images, while the discriminator learns to differentiate better between real and fake images. The better the discriminator distinguishes the images, the more the generator is pushed to generate more realistic images. This results in a so-called minimax game, where $G$ minimizes the log probability that $D$ recognizes the generated data as fake and $D$ maximizes the log probability to assign real and fake data correctly (eq.~\ref{eq:minimax}).
\begin{equation}
   \min_{G}\max_{D}V(D,G) =\mathbb{E}_{\tilde y}[\log D (\tilde y)]+\mathbb{E}_{\mathbf{z}}[\log (1-D(G(\mathbf{z})))]
\label{eq:minimax}
\end{equation}

An extension of regular GANs are the conditional GANs (cGANs), that learn the mapping from an observed image $x$ additionally,  $G:\{x,\mathbf{z}\}\rightarrow y$. 
A widespread application of cGANs is style and domain transfer ~\cite{pix2pix,pix2pixHD}, where the generator takes an image $x$ as input and is trained to generate the style-transferred image $y$ as output $G(x)\approx y$. In this case, the discriminator takes a pair of images as input and learns to determine between real pairs ($x$ and $y$) and fake pairs ($x$ and $G(x)$). In this work, the focus lies on unpaired topology-preserving domain translation, thus $x$ is an image containing edge information of $y$. This enables preserving the topology described by the edges and learning the characteristic gray value distribution of the training image domain. Hence, by simply extracting the image edges in the inference phase, it is possible to translate an image of any domain to the domain of the training data. 
\subsection{Multi-scale Patch-based Conditional GANs}
The idea of using multiple resolution scales to achieve crisper and more detailed results at the  highest resolution scale has shown to be successful in previous works. For example, in \cite{laplace}, Denton et al. use an approach based on the Laplacian pyramid by training several GANs for each pyramid level and show the efficacy of such methods. In \cite{progressive}, Karras et al. developed a new progressively growing training procedure based on training successive resolution levels iteratively and achieved impressive results. Even by using only two resolution scales, like in \cite{pix2pixHD}, much higher image resolutions are possible. The intuition behind using a multi-scale approach is that at each level a rather simple task is learned --  by taking the global image information into account, only a resolution refinement is required in each step. However, those methods assume that at a certain stage the whole full-resolution image is produced and propagated through the network. The memory demand for storing a large 3D image on the GPU is already enormous and the network propagation steps further aggravate the problem. To cope with this issue, a multi-scale patch-based approach (Fig.~\ref{fig:overview}) is proposed in this work. 

The main idea is to first generate the whole image in a very low resolution by using a low-resolution conditional GAN (LR GAN). Assume this is scale $s_0$ of the multi-resolution approach, while scale $s_n$ is the last scale at the highest resolution. To reach $s_n$ a succession of $n$ conditional high-resolution (HR) GANs is trained for each resolution level in the following manner: For HR GAN $i, i>0$, a patch from the image at scale $s_{i-1}$ serves as input. The output is a patch of the same size at scale $s_i$ representing the center of the input patch in a higher resolution. Since the patch size stays the same and the resolution grows, the receptive field of the output patch is much smaller than the input patch. For example, at level $s_0$ the image size is $64^3$ and at $s_1$ the image size increases to $128^3$. If an input patch of size $32^3$ is chosen for HR GAN 1 (Fig.~\ref{fig:overview} blue patch), then the output patch of size $32^3$ views only a sub-region (red patch) of the input since it would have the size $64^3$ at $s_1$ (green lines). In this way, the global information from scale $s_0$ is propagated to the following scales. Also by generating only a sub-region of the input low-resolution patch, each patch receives neighborhood information preventing inconsistencies at the patch borders.

To also involve the previously mentioned style transfer application, we use additional conditioning on the image edges. The LR GAN receives the low-resolution image edges of the whole image, and HR GAN~$i$ receives the edges of the patches of scale $s_i$ additionally to the low-resolution patch of scale $s_{i-1}$.   

More formally, the objective of the learning process can be expressed as in eq.~\ref{eq:objective}. For multiple conditional images $x_0\dots x_n$ with resolutions $0\dots n$, output images $y_0 \dots y_n$ are generated using separate generators $G_{0\dots n}$ and discriminators $D_{0\dots n}$ with the objectives
\begin{equation}
\begin{aligned}
    &\mathcal{L}_{cGAN}(G_0,D_0)=\\
    &\mathbb{E}_{x_0,y_0}[\log D_0(x_0,y_0)]+\mathbb{E}_{x_0,\mathbf{z}}[\log (1-D_0(x_0,G_0(x_0,\mathbf{z})))];\\
    &\mathcal{L}_{cGAN}(G_i,D_i)=\\&\mathbb{E}_{x_{p_i},y_{p_i}}[\log D_i(x_{p_i},y_{p_{i-1}},y_{p_i})]+ \mathbb{E}_{x_{p_i},\mathbf{z}}[\log (1-D_i(x_{p_i},y_{p_{i-1}},G_i(x_{p_i},y_{p_{i-1}},\mathbf{z})))].
    \end{aligned}
\label{eq:objective}
\end{equation}
Here, $x_{p_i}$ and $y_{p_i}$ are patches of the conditional image $x_i$ and the generated image $y_i$, respectively, with $i \in [1,n]$, where $y_n$ represents the final output image. To improve training stability, it is beneficial to mix the cGAN objective with a pixel-wise loss for the generator, thus we integrate the L1 distance into the objective. In our experience, leaving the pixel-wise loss out leads to a rather unstable training process since the discriminator learns much faster than the generator and thus the gradient propagated through the generator rapidly decreases, which is also consistent with \cite{pix2pix}. Also, L1-loss is preferred over L2-loss as it prevents blurriness \cite{uzunova2019unsupervised}. Further, more elaborate pixel-wise losses such as SSIM \cite{ssim} can be considered, however the performance of the L1-loss here is evaluated as sufficient and more complex losses slow down the backpropagation process significantly.
\subsection{Architectures and Training Details}
\label{sec:trainingDetails}
\begin{figure}[t]
   \centering
     \includegraphics[width=0.59\textwidth]{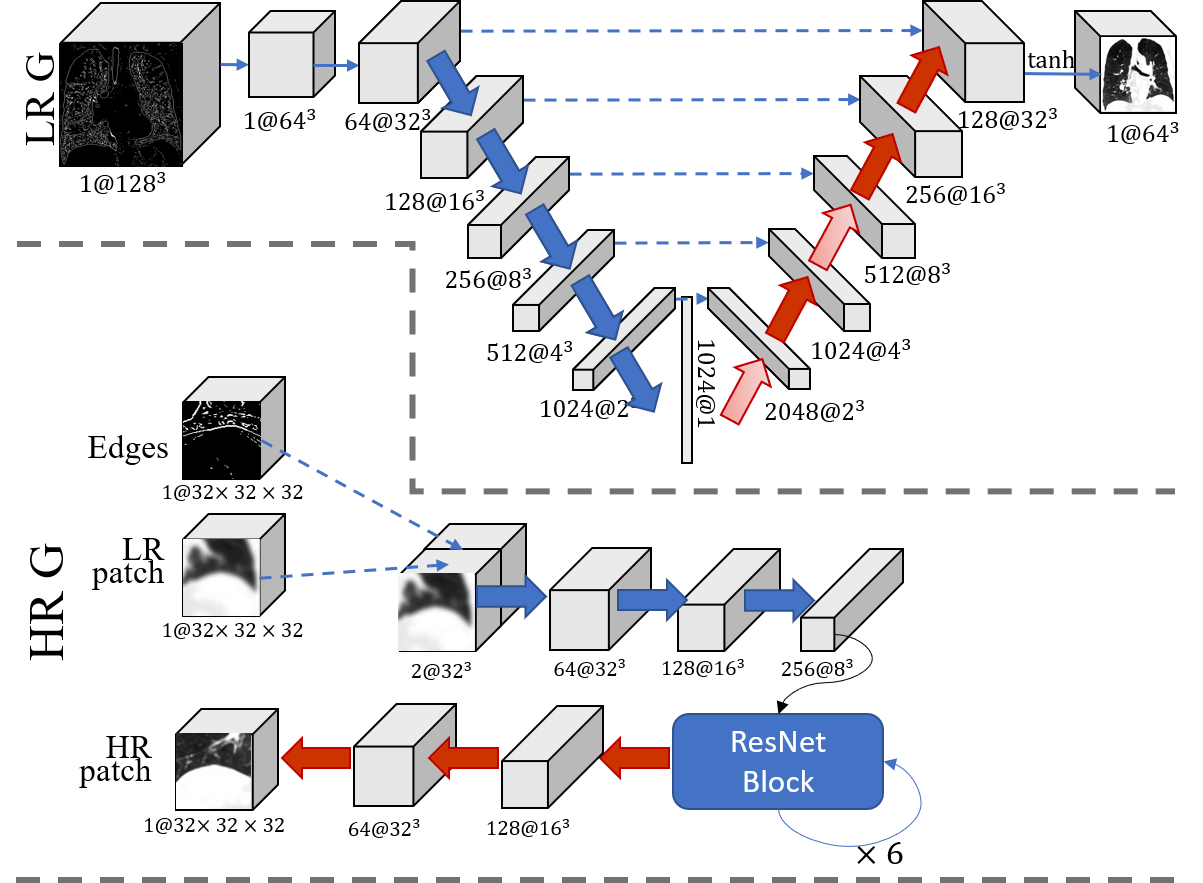}
     \includegraphics[width=0.39\textwidth]{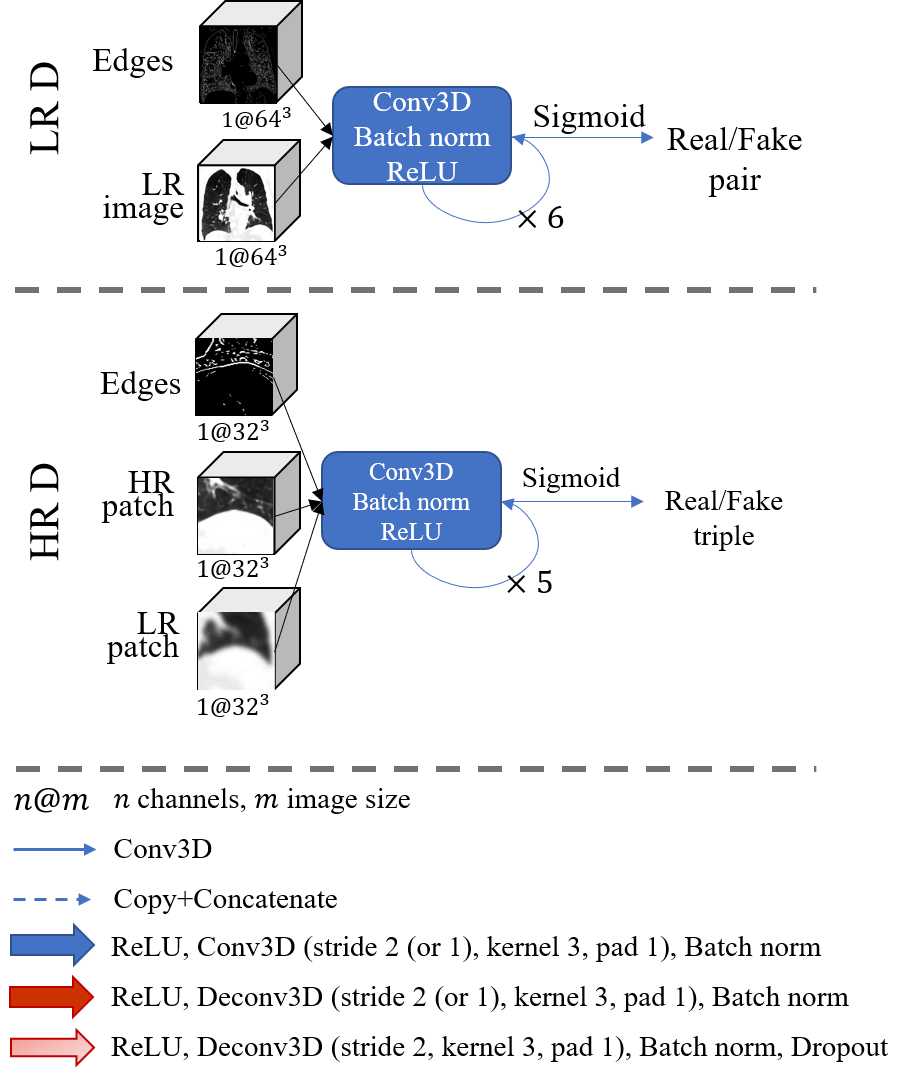}
     \caption{Generator (left) and discriminator (right) architectures.}
     \label{fig:DsGs}
\end{figure}

In an effort to demonstrate the effectiveness of the proposed method and to decouple the performance gain due to architectural search, we adopt two commonly used architectures for our generators. 
However, since the tasks of generating whole low-resolution images and high-resolution patches differ in a variety of requirements, different generator architectures were chosen for $G_0$ and $G_{1\dots n}$. The LR GAN  uses a U-Net architecture~\cite{unet}, which is able to filter out many unimportant details and generalize better due to its bottleneck. Its tendency to result in more blurry images is negligible in the context of low-resolution images. 
For the patch generation by the HR GANs, ResNet blocks~\cite{resnet} are chosen, since they are known to produce sharp results by keeping the input image resolution unchanged. The higher overfitting risk of not having a bottleneck is diminished due to the stronger conditioning (on the previous scales) and the overall large number of patches for training compared to the number of images used. For the discriminators $D_{0\dots n}$ a regular fully-convolutional architecture is chosen (s. Fig.~\ref{fig:DsGs} for detailed architectures)\footnote{Code available at: https://github.com/hristina-uzunova/MEGAN}.

Cascading approaches like these are prone to propagate errors from the lowest scale to the highest. To avoid this problem, data augmentation becomes crucial to our method. During the training of every HR generator, the low-resolution patches are augmented with noise, Gaussian blurring is applied to 30\% of the patches and the resolution of 20\% of the patches is lowered by half. The edge input patches are also distorted with noise. These measures hinder overadapting to each of the inputs and help to cope with imperfections in the edge extraction method and the generated images of lower scales at inference phase. Also, the augmentation with noise and the usage of dropout replace the straight-forward input noise vector $\mathbf{z}$, since in that case it would simply get ignored~\cite{noise}.

Further, the edge image used as input for the LR GAN has twice the resolution of the output LR image to prevent the loss of too much edge information. Also, in our experience, it showed to be advantageous not to use strictly binary edges, but gradient magnitude weighted Canny-extracted edges.

In order to avoid padding-related patch artifacts, only the network's receptive field from each patch is used when reconstructing the generated images, as proposed in \cite{deepmedic}.

In all experiments, an image size of $64^3$ is produced by the LR GAN (input edge image of size $128^3$) and iteratively upscaled by doubling the resolution with HR GANs producing patches of size $32^3$ until the original image size is reached.  The choice of the image size for the LR GAN is of significant importance, since a trade-off between capturing all important global information of the images and avoiding too many details needs to be made. Thus an LR size of $32^3$ is not sufficient to represent the complexity of the images, while $128^3$ contains too many details that cannot be captured by the LR GAN. Naturally, the input image size needs to be adjusted depending on the size and complexity of the available data. Using larger patches for the HR GAN does not introduce significant improvements, however more computational resources are demanded.

\section{Experiments and Results}

\subsection{Memory Requirements for 3D Images}
\label{sec:memreq}
 GANs are currently rarely applied to 3D images due to computational constraints, therefore in this experiment, the dependence of 3D image size and memory requirements of different GAN-based image generation techniques is investigated. Three common GAN architectures are chosen as baselines: DCGAN~\cite{3dgan}, Pix2Pix~\cite{pix2pix} and progressive growing GAN (PGGAN)~\cite{progressive}, and compared to the two architectures of our method: LR 64 for low-resolution images of size $64^3$  and HR 32 for high-resolution patches of size $32^3$.
 PyTorch~\cite{pytorch} is used as an implementation framework of choice for all networks.
 Since Pix2Pix and PGGAN are only implemented for 2D images, a straight-forward translation to 3D is obtained (replacing 2D by 3D convolutions, etc.). The RAM demand computation is realized using an approach similar to the summary approach from the keras framework~\cite{keras}. The assumed lower bound of memory usage here includes one forward and backward pass for the generator and discriminator each, as well as the memory required to store the images, gradients and network parameters for batch size one.
 
\begin{figure}[h!]
\centering
     \includegraphics[width=0.7\textwidth]{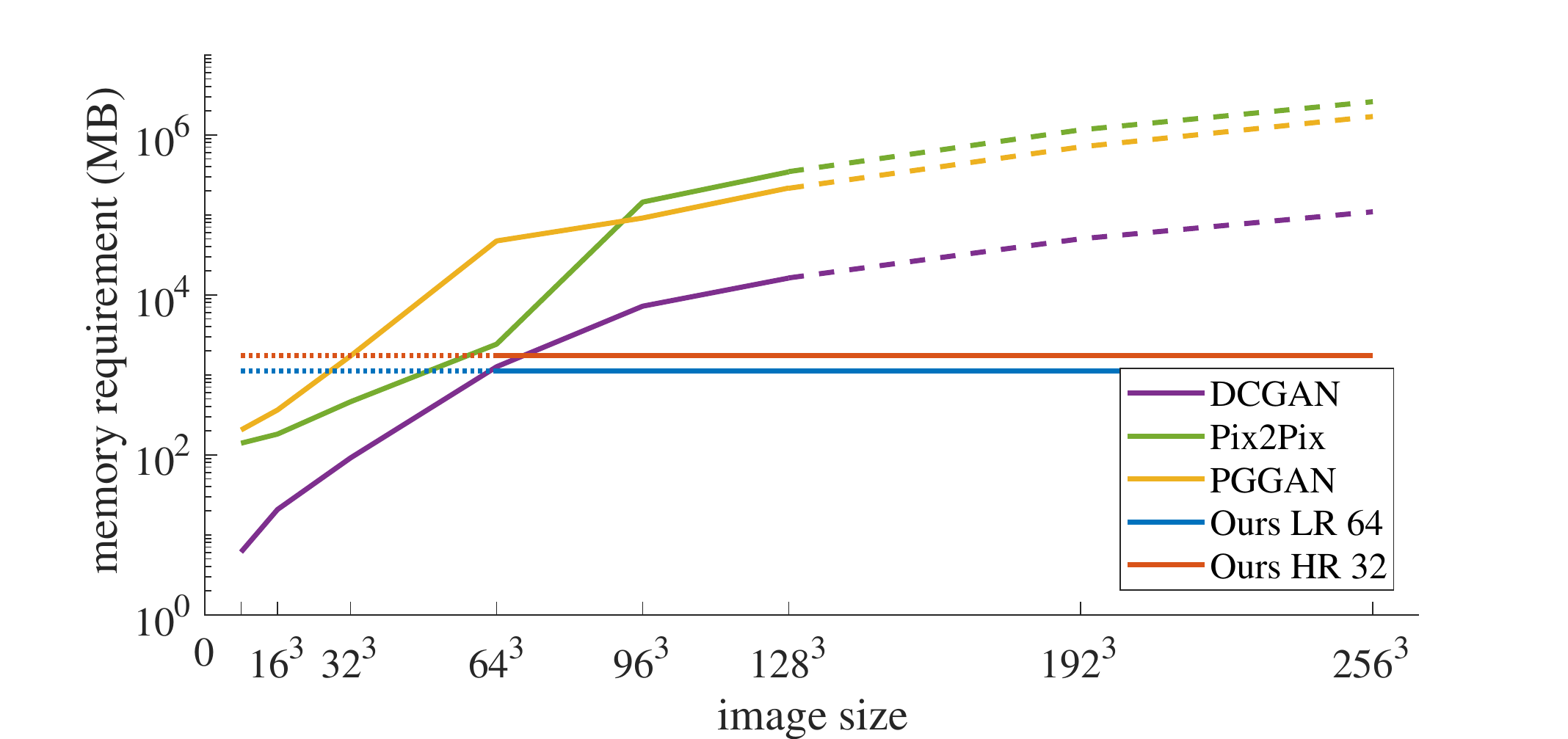}
     \caption{RAM requirements for 3D GANs. Baselines: DCGAN,  Pix2Pix and PGGAN. Dashed lines indicate cubic regression approximation. Our methods: for low resolution images of size $64^3$ (LR 64) and high resolution patches of size $32^3$ (HR 32), have constant memory requirement regardless the image size. Dotted lines indicate sizes under the assumed minimal size  $64^3$. Log-scale is used on the $y$-axis.}
     \label{fig:memuse}
\end{figure}

 The results for different image sizes are shown in Fig.~\ref{fig:memuse}. Naturally, all three baseline approaches have at least cubic memory requirement growth w.r.t. the image side's length. For those approaches calculations for size over $128^3$ were not even possible on the used Titan XP 12GB GPU, thus the extrapolated cubic regressed curves are shown. These results underline the infeasibility of straight-forward 3D GAN approaches for medical images, as their sizes commonly reach $512^3$, e.g., PGGAN require more than $100$GB GPU RAM for images of size $256^3$. In contrast, our method is of constant nature w.r.t. the image side length and is thus suitable for arbitrary image sizes with predictable memory usage.      
  
\subsection{Domain Translation for Medical Images}
\label{sec:domaintranslation}
 Usually in medical imaging, there is a large variety of different acquisition parameters that can be chosen to ensure different contrast and better visibility of particular tissues. This flexibility, however, yields inconsistent acquisitions between datasets, causing inconvenience for automated data analysis algorithms. One example are the different reconstruction kernels used in CT imaging. While sharp kernels are more pleasant for visual perception, soft kernels are more appropriate for automatic calculations. Another common problem occurs in MRI imaging, since different pulse sequences can be chosen (e.g. T1-,T2-weighted, FLAIR). Many algorithms require complete datasets or specific sequences (e.g. segmentation with FreeSurfer \cite{freesurfer}), however, in many cases not all of the sequences are available. For those reasons, medical image domain translation for the generation of missing image modalities becomes crucial for automatic image analysis applications. And although, there are many domain translation frameworks, most of them need paired data for training and do not enable large image sizes. In our experiments, we apply the proposed method for unpaired domain translation of high resolution 3D medical images. However, domain translated synthetic images for direct diagnostic purposes by a clinician is not implied, still, we suggest their application in many different image analysis algorithms.    


\subsubsection{Domain Translation for 3D Thorax CTs}
For those experiments the dataset \emph{Lungs COPD} is used for training. \emph{Lungs COPD} is a thorax CT dataset, containing 3D CT images of size ca. $512^3$ of 56 subjects with varying degree of chronic obstructive pulmonary disease (COPD)\cite{lungsCOPD}. For each subject the data was simultaneously reconstructed with different kernels: soft (B20f), sharp (B50f) and very sharp (B80f). 
In the following, one LR and three HR GANs are trained  in a 5-fold-cross-validation manner on the B20f images from the  dataset. Those experiments aim to show the possibility of our method to generate very large (up to 512$^3$) images in high resolution and quality. 

\begin{figure}[h!]
\centering
\begin{tabular}{c@{\hspace{-0pt}}c@{\hspace{-0pt}}c@{\hspace{-0pt}}c@{\hspace{-0pt}}}
\includegraphics[width=0.2\textwidth]{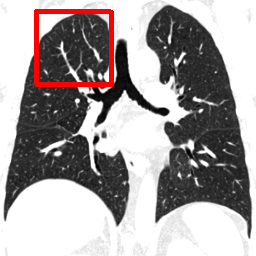}&
\includegraphics[width=0.2\textwidth]{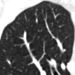}&
\includegraphics[width=0.2\textwidth]{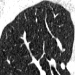} &
\includegraphics[width=0.2\textwidth]{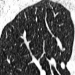} \\
\small{B20f} & \small{B20f (target)} & \small{B50f} & \small{B80f} \\
\includegraphics[width=0.2\textwidth]{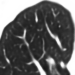} &
\includegraphics[width=0.2\textwidth]{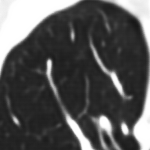} &
\includegraphics[width=0.2\textwidth]{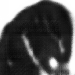}&
\includegraphics[width=0.2\textwidth]{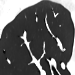} \\
\small{Our method} & \small{Our method} & \small{Patch} & \small{NLM}\\
\small{B50f$\rightarrow$B20f}&\multicolumn{3}{c}{\small{B80f$\rightarrow$B20f}}  \\
\includegraphics[width=0.2\textwidth]{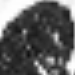} &
\includegraphics[width=0.2\textwidth]{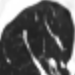} &
\includegraphics[width=0.2\textwidth]{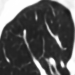}&
\includegraphics[width=0.2\textwidth]{images/fake_level3_G6_G6_B80f_cut.png} \\
\small{Scale 0: $64^3$} & \small{Scale 1: $128^3$} & \small{Scale 2: $256^3$} & \small{Scale 3: $512^3$}\\
\multicolumn{4}{c}{\small{B80f$\rightarrow$B20f}} 
\end{tabular}
\caption{Sharp (B80f and B50f) to soft (B20f) kernels. From left to right: First row -- original images: Representing real B20f axial coronal slice from a 3D thorax CT (red: zoom-in area for all other images); Real B20f; Real B50f; Real B80f. Second row -- generated B20f images by: Our method from the B50f images; Our method from the B80f images; A straight-forward patch-wise approach from the B80f images; NLM smoothed B80f images. Third row -- different scales of our method. }
\label{fig:lungs}
\end{figure}

\paragraph*{Sharp to Soft Kernels} 
The influence of different CT acquisition parameters on the image quality usually hinders the comparability of automatic image quantifications between various settings, e.g. the emphysema index in CTs reconstructed with different kernels \cite{emphysema1,emphysema2}.
  For this reason we obtain domain translations from sharper kernels (B80f and B50f) of the \emph{Lungs COPD} to the soft kernel (B20f) using the presented GAN-based method.
 In the inference phase, the edges of the B80f and B50f image from the current test fold are extracted and propagated through the cascading nets. This results in keeping the topology of the noisy images and filling them with gray values typical for the smooth B20f images. Examples of the domain translation results can be seen in Fig.~\ref{fig:lungs}. 

Since in this dataset every image is simultaneously reconstructed with each kernel, the corresponding ground truth B20f image for each B80f and B50f reconstruction is available, and thus a quantitative comparison is enabled. The calculated measurements between the generated and ground truth images are the structured self-similarity index (SSIM)~\cite{ssim}, peak signal-to-noise ratio (PSNR), mean average error (MAE), mean squared error (MSE). Note, that the B20f reconstructions of the test images are used for evaluation only and are whether requested for training, nor inference. Also, only one training on the B20f data is sufficient for translations from every domain (here: B80f and B50f). 

For comparison the following standard approaches are implemented: 1) Straight-forward patch-wise GAN -- a GAN trained in the exact same manner as the HR GAN on the last scale, except for using the information from the lower resolution scales. In the test phase, patch overlaps are applied to avoid artifacts. 2) Straight-forward GAN for smaller image size and then rescaling to the original size. For this method, the LR GAN architecture is applied to generate images of size $64^3$ (this is on the edge of our computational abilities), followed by trilinear interpolation. 3) Smooth the B80f/B50f images to achieve the appearance of a B20f image. This is accomplished using the state-of-the-art edge-preserving non-local means filter (NLM). 

The quantitative results for all methods and both scenarios (B80f$\rightarrow$B20f and B50f$\rightarrow$B20f) are shown in tab.~\ref{tab:resultsLungs}. After domain translation of the B80f images with our method, the images are significantly more similar to the ground truth, especially when SSIM is taken into account. However, since the B50f and B20f images are initially more similar, the B50f$\rightarrow$B20f translation, does not deliver such large discrepancy in values, however, the SSIM of the translated images is still significantly higher compared to the original ones, suggesting a good domain adaption. Also, the results are visually appealing (Fig.~\ref{fig:lungs}). Even though the NLM smoothed images show similar quantitative results, qualitatively they lack many important details and the filtering requires $\sim$ 100 times longer computational time. The patch-based and resizing methods deliver images of significantly worse quality both visually and quantitatively. To illustrate the importance of using multiple scales, the results of four scales are visualized in Fig.~\ref{fig:lungs}, where there is a clear increase in quality  with each scale.

\begin{figure}[h]
\centering
 \begin{tabular}{c@{\hspace{0.1pt}}c@{\hspace{0.1pt}}c@{\hspace{0.1pt}}c@{\hspace{0.1pt}}}
 \centering
\includegraphics[width=0.2\textwidth]{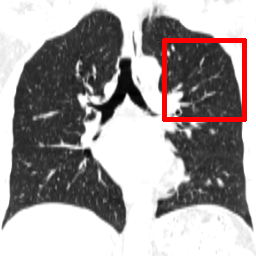} &
\includegraphics[width=0.2\textwidth]{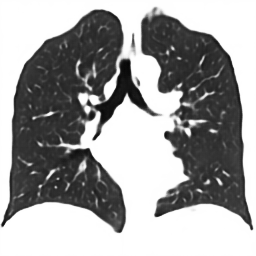}&
\includegraphics[width=0.2\textwidth]{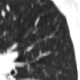} &
\includegraphics[width=0.2\textwidth]{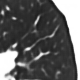} \\
\small{Low-dose} & \small{High-dose (fake)} & \small{Low-dose} & \small{High-dose (fake)}\\
\end{tabular}
\caption{Low-dose to high-dose domain translation example image. Left two: Whole images; Right two: Zoomed-in into the red square.}
\label{fig:lowtohigh}
\end{figure}

\begin{table}[h]
\centering
\caption{Quantitative results for the lungs CT experiments.Measurements between a generated image and its ground truth. Columns 3-6: average SSIM (higher is better), MAE and MSE (lower is better), PSNR (higher is better). Note that the image intensities are normalized within [0,1]. Experiments (top and bottom): B80f to B20f image translation, B50f to B20f image translation. Compared to ground truth (row-wise):  our generated images, conventional patch-wise generation, up-scaled low-resolution images, a non-local means filtered image \cite{nlm} and the original image. Superscripts correspond to significance ($p<0.0001$) in a paired two-sided t-test for all methods compared to ours in terms of: all measures $^\star$; SSIM $^\dagger$.}
\label{tab:resultsLungs}
\footnotesize

\begin{tabular}{l@{\hspace{0.5em}}| l@{\hspace{0.5em}}c@{\hspace{1em}}c@{\hspace{1em}}c@{\hspace{1em}}c}
\footnotesize
               & & \textbf{SSIM} & \textbf{MAE} & \textbf{MSE}& \textbf{PSNR (dB)}\\
\textbf{}& \textbf{Method} & mean & mean & mean & mean
\\\hline
\multirow{5}{*}{B80f$\rightarrow$B20f} 
                            & Our gen.&${0.773}$&$0.033$&${0.004}$ & $24.1$\\
                            & Patch gen.${^\star}$ &$0.706$&$0.049$&$0.008$ &$21.2$\\
                            & Small gen.${^\star}$ &$0.633$&$0.058$&$0.011$ &$19.5$\\
                            & NLM B80f&${0.773}$&${0.031}$& ${0.004}$ & $19.5$\\
                            & Orig. B80f${^\star}$ &$0.480$&$0.065$&$0.012$ & $19.2$\\
\hline
\multirow{5}{*}{B50f$\rightarrow$B20f}
                            & Our gen.&${0.794}$&$0.033$&$0.003$&$24.8$\\
                            & Patch gen.${^\star}$   &$0.698$&$0.050$&$0.008$&$21.1$\\
                            & Small gen.${^\star}$  &$0.636$&$0.055$&$0.012$&$19.3$\\
                            & NLM B50f${^\star}$  &$0.478$&$0.283$& $0.213$&$19.3$\\
                            & Orig. B50f${^\dagger}$  &$0.722$&$0.028$&$0.002$&$26.3$\\
\hline
\end{tabular}

\end{table}
\paragraph*{Low-dose to High-dose} This experiment underlines the ability of the presented method for unpaired domain translation. The domain translation is established from a completely different dataset -- the \emph{Low-dose Lungs} --  to the B20f \emph{Lungs COPD} domain with a simple inference through the network trained in the previous experiment. The \emph{Low-dose Lungs} dataset contains 10 3D inhale phase thorax CTs from the COPDgene dataset~\cite{copdgene} of sizes around $300^3$ all having COPD. Since the  \emph{Lungs COPD} B20f training data are acquired with a higher dose than the \emph{Low-dose Lungs} data, this experiments corresponds to a low-dose to high-dose translation. The edges of all ten images from the low-dose dataset are extracted and propagated through the trained GANs. Visually the generated images look smoother and of better quality (Fig.~\ref{fig:lowtohigh}) than the original low-dose images. Due to the lack of ground truth for this experiment (no corresponding high-dose images are available), only visual evaluation is possible. Note, that those experiments do not aim the generation of data for direct interpretation (by clinicians), since GANs tend to hallucinate pathologies and other features (here: emphysema)~\cite{hallucinate}, however, automatic image analysis methods can be facilitated.

\subsubsection{Domain Translation and Data Augmentation for 3D Brain MRIs}
The following experiments are based on brain MRI domain translation using the \emph{BRATS} dataset for training. This dataset features 274 multi-modal 3D brain MRI images from the BRATS challenge~\cite{brats}. All images contain brain tumorous of the type glioblastomas with given expert segmentations. The initial image size is $155\times240\times240$, however the images are cropped with a bounding box around the brain. In this work the Flair, T1- and T2-weighted sequences are considered, where the T2-weighted image sequences are used in a 4-fold-cross-validation manner. One LR and two HR GANs are trained on the data. Since the segmentation masks of the tumors are given, they get strictly integrated into the training and testing process by overlaying them on the extracted edges. In our experience, the explicit integration of the tumors (and not only their extracted edges), help avoid pathology hallucination~\cite{hallucinate}. In the following experiments, the aim is to explore the boundaries of the proposed method and its separate components.
\paragraph*{MRI Modality Translation} The task of translating between different MRI modalities is quite often requested, thus this experiment shows that our method is also suitable for MRI modality translation on the full-sized 3D images from the \emph{BRATS} dataset. For inference the edges from the Flair as well as T1-weighted sequences of the test fold are cascaded through the trained networks, resulting into a Flair$\rightarrow$T2 and a T1$\rightarrow$T2 translation. In this dataset, all three sequences for each image are available and co-aligned. Hence a ground truth for quantitative evaluation is available. 

As a comparison, two state-of-the art paired image translation methods are considered and trained on the same folds: the random forest based REPLICA\cite{REPLICA} method; and the patch-wise 3D GAN MedSynth\cite{medsynth} approach  that applies an auto-context model for patch refinement. For both methods, the source code provided by the authors was adapted for the experiments. The results can be seen in tab.~\ref{tab:resultsBrains}. The overall high SSIM and PSNR, and low MAE and MSE indicate good correspondence between the generated and real T2-weighted images. All methods deliver comparable results in terms of MSE and MAE, where as expected paired methods tend to generate intensities more similar to the original images. However, in terms of SSIM, describing the structural information of the images, the presented method delivers significantly better results, capturing its ability to generate sharper images and no visible patch artifacts. The low SSIM values of the MedSynth images can be explained by the often visible patch and slice artifacts of the generated images, especially in tumor regions. 

\begin{table}[t]
\centering
\caption{Quantitative results of the Brain MRI experiments.
Measurements between a generated image and its ground truth. Columns 3-6: average SSIM (higher is better), MAE and MSE (lower is better), PSNR (higher is better). Note that the image intensities are normalized within [0,1]. Experiments (top and bottom): T1 to T2 translation, Flair to T2 translation. Compared to ground truth (row-wise) generated images by our method, REPLICA and MedSynth. Superscripts correspond to significance ($p<0.0001$) in a paired two-sided t-test for all methods compared to ours in terms of SSIM $^\star$.}
\label{tab:resultsBrains}
\footnotesize
\begin{tabular}{l@{\hspace{0.5em}}| l@{\hspace{0.5em}}c@{\hspace{1em}}c@{\hspace{1em}}c@{\hspace{1em}}c}
\footnotesize
               & & \textbf{SSIM} & \textbf{MAE} & \textbf{MSE}& \textbf{PSNR (dB)}\\
\textbf{}& \textbf{Method} & mean & mean & mean & mean
\\\hline
\multirow{3}{*}{T1$\rightarrow$T2} 
                            & Ours$^\star$&$0.911$&$0.017$&$0.003$ & $26.0$\\
                            & REPLICA \cite{REPLICA} &$0.854$&$0.017$&$0.003$ &$26.9$\\
                            & MedSynth \cite{medsynth} & $0.613$ & $0.025$ &$0.003$&$27.0$\\
\hline
\multirow{3}{*}{Flair$\rightarrow$T2}
                            & Ours$^\star$&$0.905$&$0.021$&$0.004$ & $24.6$\\
                           & REPLICA \cite{REPLICA}&$0.833$&$0.019$&$0.002$ &$25.9$\\
                           & MedSynth \cite{medsynth}& $0.734$ & $0.020$ &$0.002$&$27.5$\\
\hline
\end{tabular}
\end{table}

\begin{figure}[h]
\centering
 \begin{tabular}{c@{\hspace{1em}}c@{\hspace{1em}}c@{\hspace{1em}}c@{\hspace{1em}}c@{\hspace{1em}}}
 \centering
\includegraphics[width=0.15\textwidth,trim={1.5cm 1cm 4cm 1.4cm},clip]{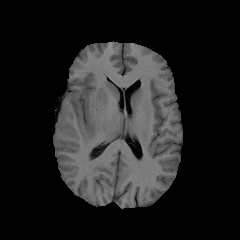} &
\includegraphics[width=0.15\textwidth,trim={1.5cm 1cm 4cm 1.4cm},clip]{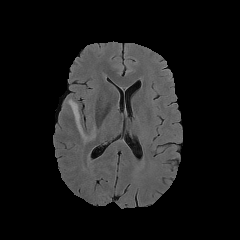}&
\includegraphics[width=0.15\textwidth,trim={1.5cm 1cm 4cm 1.4cm},clip]{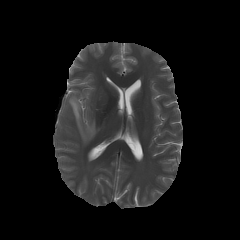} &
\includegraphics[width=0.15\textwidth,trim={1.5cm 1cm 4cm 1.4cm},clip]{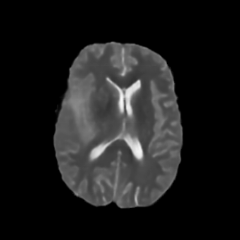}&
\includegraphics[width=0.15\textwidth,trim={1.5cm 1cm 4cm 1.4cm},clip]{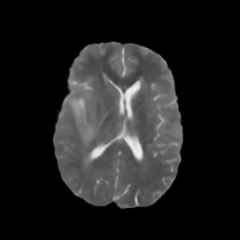}\\
\includegraphics[width=0.15\textwidth,trim={1.5cm 1cm 4cm 1.4cm},clip]{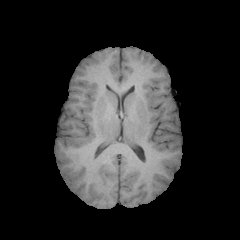} &
\includegraphics[width=0.15\textwidth,trim={1.5cm 1cm 4cm 1.4cm},clip]{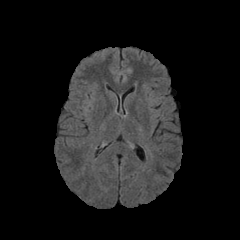}&
\includegraphics[width=0.15\textwidth,trim={1.5cm 1cm 4cm 1.4cm},clip]{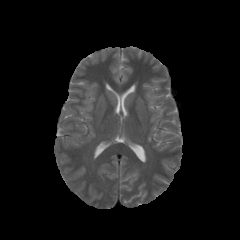} &
\includegraphics[width=0.15\textwidth,trim={1.5cm 1cm 4cm 1.4cm},clip]{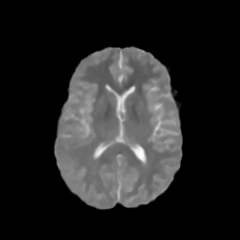}&
\includegraphics[width=0.15\textwidth,trim={1.5cm 1cm 4cm 1.4cm},clip]{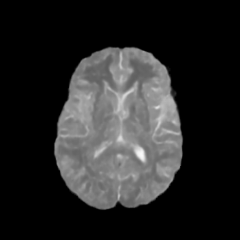}\\
\small{Real T1}&\small{Real Flair}&\small{Real T2}&\small{T1$\rightarrow$T2}&\small{Flair$\rightarrow$T2}\\
\end{tabular}
\caption{MRI sequence translation (T1$\rightarrow$T2) and (Flair$\rightarrow$T2). Row-wise: slices from two example BRATS images in their three sequences (left three) and the domain translated images generated by our method (right two), cropped to the left brain half for better visibility.}
\label{fig:t12t2}
\end{figure}
Examples of domain translated images of our method can be seen in Fig.~\ref{fig:t12t2}. Visually, the generated images have a realistic appearance, however, the translation from the T1-weighted sequence delivers sharper visual and better quantitative results than from Flair, since the reduced contrast of the Flair images impedes edge extraction. Even though the results are satisfactory, the dependence of the presented method on well-extracted edges gets highlighted. 

\begin{figure}[h]
\centering
 \begin{tabular}{c@{\hspace{0.1pt}}c@{\hspace{1em}} c@{\hspace{0.1pt}}c@{\hspace{0.1pt}}}
 \centering
\includegraphics[width=0.2\textwidth,trim={2.8cm 1.8cm 3.8cm 4.8cm},clip]{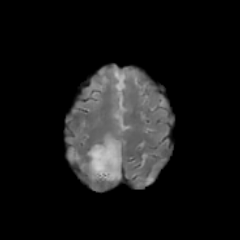} &
\includegraphics[width=0.2\textwidth,trim={2.8cm 1.8cm 3.8cm 4.8cm},clip]{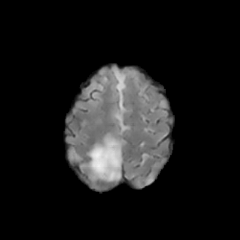}&
\includegraphics[width=0.2\textwidth,trim={4.5cm 3cm 1.5cm 3cm},clip]{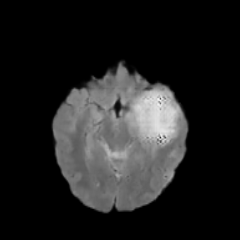} &
\includegraphics[width=0.2\textwidth,trim={4.5cm 3cm 1.5cm 3cm},clip]{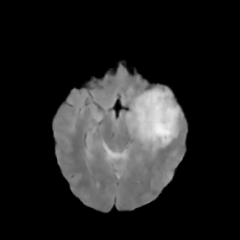}\\
scale 2 & scale 3 & scale 2 & scale 3
\end{tabular}
\caption{Zoom-ins on the tumor tissue from scale 2 and 3 of two generated images. Artifacts generated on lower scales do not get propagated to the high resolution images.}
\label{fig:artifacts}
\end{figure}

Since the generation of realistically looking tumor tissue in MR images is a quite complicated task, the amount of artifacts generated in the tumor tissue, especially on the lower scales, is considerable. However, thanks to the augmentation approaches mentioned in Sec.~\ref{sec:trainingDetails}, the artifacts from lower scales, in general do not get propagated to the last scale (s. Fig.~\ref{fig:artifacts} for examples).

\begin{figure}[t]
\centering
\begin{tabular}{c@{\hspace{1em}}c@{\hspace{1em}}c@{\hspace{1em}}}
\includegraphics[height=0.2\textwidth]{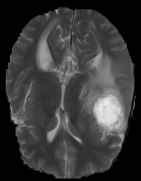}&
\includegraphics[height=0.2\textwidth]{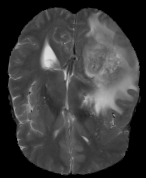}&
\includegraphics[height=0.2\textwidth]{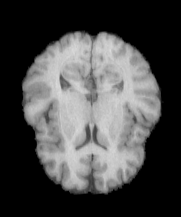}\\
\small{BRATS T2 1} & \small{BRATS T2 2} & \small{LPBA} \\
 \end{tabular}

 \begin{tabular}{c@{\hspace{1em}}c@{\hspace{1em}}c@{\hspace{1em}}c@{\hspace{1em}}}
 \centering
\includegraphics[width=0.2\textwidth]{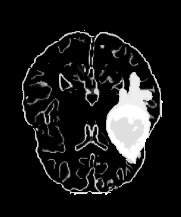} &
\includegraphics[width=0.2\textwidth]{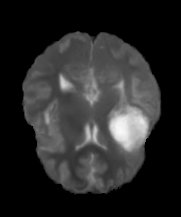}&
\includegraphics[width=0.2\textwidth]{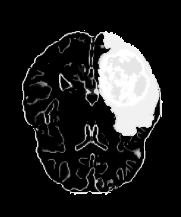} &
\includegraphics[width=0.2\textwidth]{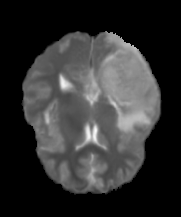} \\
\small{Sketch LPBA}  & \small{Fake T2} & \small{Sketch LPBA}& \small{Fake T2}\\
\small{+ tumor 1} & \small{+ tumor 1} &  \small{+ tumor 2} &  \small{+ tumor 2} 
\end{tabular}
\caption{LPBA to BRATS T2 domain translation examples. First row: Real images; Second row: Extracted edges and generated (fake) images.}
\label{fig:lpba2brats} 
\end{figure}

\begin{figure}[h!]
\centering
 \begin{tabular}{c@{\hspace{1em}}c@{\hspace{1em}}c@{\hspace{1em}}c@{\hspace{1em}}}
 \centering
\includegraphics[width=0.2\textwidth]{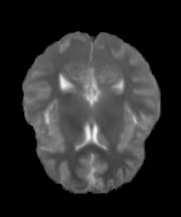}&
\includegraphics[width=0.2\textwidth]{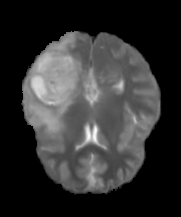}&
\includegraphics[width=0.2\textwidth]{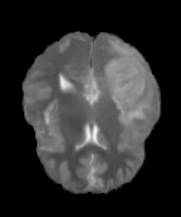} &
\includegraphics[width=0.2\textwidth]{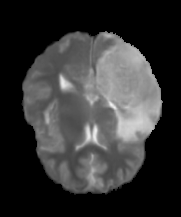} \\
&
\includegraphics[width=0.2\textwidth]{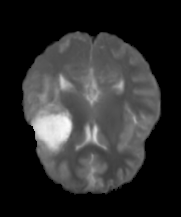}&
\includegraphics[width=0.2\textwidth]{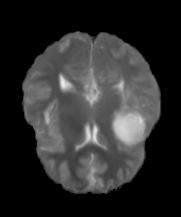} &
\includegraphics[width=0.2\textwidth]{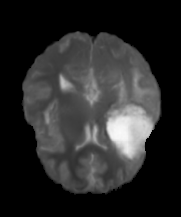}\\
\small{No tumor}  & \small{Mirror tumor} & \small{Shrink tumor}& \small{Zoom tumor}\\
\end{tabular}
\caption{Tumor augmentation examples by applying affine transformations of the tumor label. Row-wise: Two different tumors applied on the same image (tumor 1 and tumor 2 from Fig.~\ref{fig:lpba2brats}). }
\label{fig:tumoraug}
\end{figure}

\paragraph*{Healthy to Pathological Translation} To once again underline the ability of our method for unpaired domain translation, here, a domain translation between the healthy patients T1-weighted MRIs from a completely different dataset (\emph{LPBA}) and the pathological T2-weighted \emph{BRATS} dataset is established. The LPBA dataset~\cite{lpba40} contains 40 healthy T1-weighted MRIs of size $181\times217\times 217$. Since the edges of the \emph{LPBA} images contain no pathologies, the \emph{BRATS} domain cannot be completely obtained. However, the explicit integration of the tumor mask allows to overlay the mask on the \emph{LPBA} edges and thus generate pathological data (Fig.~\ref{fig:lpba2brats}). It can also be observed that leaving the tumor mask out, does not lead to hallucination of pathologies, even though the training dataset is entirely pathological (Fig.~\ref{fig:tumoraug}). 

A further advantage of the explicit tumor mask integration is the possibility to use it for data augmentation or dataset balancing purposes, e.g., when having only a few pathological cases in a training dataset. The tumor masks can be transformed in simple ways creating new appearances of the tumors. In our experiments, the transformations feature shrinking and zooming by 15\% and mirroring the tumor on the y-axis (Fig.~\ref{fig:tumoraug}). Even though the generated images look highly realistic, they lack some essential medical details such as tissue compression around the tumor. This is in agreement with \cite{hallucinate}, that artificial data generated by GANs should not be used for direct interpretation by clinicians, but can be very helpful for automatic image analysis algorithms. 

\begin{figure}[h!]
\centering
 \begin{tabular}{c@{\hspace{1em}}c@{\hspace{1em}}c@{\hspace{1em}}c@{\hspace{1em}}}
 \centering
\includegraphics[width=0.2\textwidth]{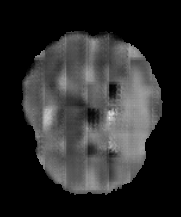} &
\includegraphics[width=0.2\textwidth]{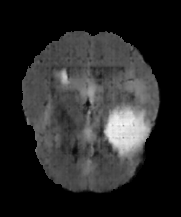}&
\includegraphics[width=0.2\textwidth]{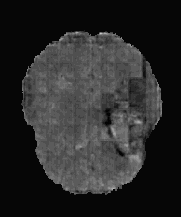} &
\includegraphics[width=0.2\textwidth]{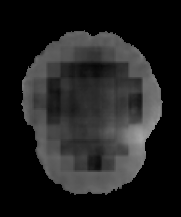} \\
\small{Only edges}&\small{Only multi-scale}&\small{Only edges}&\small{Only multi-scale}\\
\small{Train}&\small{Train}&\small{Test}&\small{Test}\\
\end{tabular}
\caption{Edges vs. multi-scale: examples of a generated image when leaving out one of the inputs (fake T2 + tumor1 from Fig.~\ref{fig:lpba2brats}). Two images on the left: One input is left out during training and testing; Two images on the right: One input is left out during testing only.}
\label{fig:edges}
\end{figure}

\paragraph*{Edges vs. Multi-scale} The intuition behind using both multiple resolution scales and edge information is to provide every HR GAN  with fine information from the edges and coarse gray value information from the previous resolution scale (similar to a Laplacian pyramid). To test this assumption, two experiments are carried out. First, the HR GANs are trained again (the LR GAN stays unchanged) for the LPBA$\rightarrow$BRATS T2 translation scenario: once without considering edge information and once without using the images from previous scales as input. While this experiment shows the importance of each component during training, the second experiment considers both components for training and shows the effect of each of them in the test phase. Here, the HR GANs trained on both inputs are used, and in the inference phase, one of the inputs consecutively gets blacked out. Example results can be seen in Fig.~\ref{fig:edges}. Surprisingly, training the network on only one of the inputs significantly reduces the quality of the generated images. When using only edge information for training, inconsistent appearance can be achieved and patch artifacts are strongly visible. As opposed to that, using a sole multi-scale approach for training generates blurry images due to the lack of high-resolution information. When simply leaving one of the inputs out during testing, using only the edges creates a rather average gray-value profile and a patchy appearance. Interestingly, when the edge information is left out, the inner part of the brain barely gets reproduced, supposedly due to the large amount of details in this area.

\section{Conclusion}
\label{sec:conclusion}
In this work, we propose a multi-scale GAN-based approach for the memory-efficient generation of high-resolution 3D images. The leading idea of the method is to first generate the whole image in a very low resolution using an LR GAN, followed by a succession of HR GANs that generate patches of the same size but growing resolutions. Since the patches are conditioned on previous scales, no patch artifacts appear. In this way, the memory demand remains constant independently from the image size, allowing for arbitrarily large image generation. The multi-scale scheme also allows creating images of very high quality and resolution, compared to conventional GAN-based methods. Due to the augmentation techniques applied during training a typical problem of propagating  errors from lower scales to higher scales is avoided. The experiments in this work are based on a domain translation scenario. Different translation tasks on a thorax CT (sharp to soft kernels, low-dose to high-dose) and a brain MRI (T1/Flair to T2, healthy to pathological) dataset were investigated and demonstrate the wide range of applications for the presented methods. This is established by additionally conditioning the GANs on an edge based image sketch. Alongside with the suitability of the method for those domain translation tasks, the experiments also show the importance of using a combination of the image edges and a multi-scale approach, so that local high-resolution and global low-resolution information is combined. Compared to paired methods, the proposed approach delivers comparable results in quality but carries significant advantages such as explicit pathology integration (and thus less feature hallucination) and augmentation. This opens up possibilities for data augmentation, dataset balancing, and various preprocessing steps for automatic algorithms, however, the images should not be considered for direct usage by clinicians due to the high probability of feature hallucination. 

However, the experiments also highlight a drawback of using the edge information as a topology constrain, since imprecise edge extraction leads to worse quality of the generated images. This would impair tasks like CT to MRI image translation, as their extracted edges lack correspondence. A more elaborate topology constrain is required to solve this issue and will be investigated in our future work.

\section*{Acknowledgements}
This research did not receive any specific grant from funding agencies in the public, commercial, or not-for-profit sectors.
\bibliography{mybibfile}

\end{document}